\documentclass[aps,prl,preprint]{revtex4}

\input{psfig.sty}

\begin{document}

\draft

\title
{\large\bf Local Distortion Induced Metal-to-Insulator Phase Transition
in PrRu$_4$P$_{12}$.}

\author{D. Cao,$^1$ R.~H. Heffner,$^1$ F. Bridges,$^{2}$ I.-K. Jeong,$^1$
E.~D. Bauer,$^1$ W.~M. Yuhasz,$^3$ and M.~B. Maple,$^3$}

\address{$^1$MS-K764, Los Alamos National Laboratory, Los Alamos, NM 87545}
\address{$^2$Physics Department, University of California, Santa Cruz, CA 95064}
\address{$^3$Department of Physics and Institute for Pure and Applied Physical Science, University of California, San Diego, LaJolla, CA 92093}

\date{\today}

\begin{abstract}

Extended X-ray Absorption Fine Structure (EXAFS) experiments have been carried
out on PrRu$_4$P$_{12}$ and PrOs$_4$P$_{12}$ to study the metal-to-insulator
(MI) phase transition in PrRu$_4$P$_{12}$. No Pr displacement was observed
across the MI transition temperature from the EXAFS data. Instead, our EXAFS
data clearly show that a Ru displacement is associated with this MI transition.
The very high Debye temperature for the Ru-P bond ($\Theta_D$=690~K) suggests
that a slight rotation/displacement of relatively rigid RuP$_6$ octahedra leads
to this small Ru displacement, which accompanies the MI transition at 62~K in
PrRu$_4$P$_{12}$.

\end{abstract}

\maketitle

The filled skutterudites have generated significant interest due to a broad
range of novel physical properties, which include unconventional
superconductivity, heavy fermion, and non-Fermi liquid behavior, as well as
their potential thermoelectric applications in industry. Transport measurements
show that most of these materials are metals with a very poor thermal
conductivity, with several phosphide skutterudites (such as CeFe$_4$P$_{12}$
and CeRu$_4$P$_{12}$) being semiconductors instead.
\cite{Meisner85,Shirotani99} Despite theoretical attempts to calculate the band
structure of the filled skutterudites, a good understanding of the
semiconducting behavior in these phosphides is still lacking. 

Recently, PrRu$_4$P$_{12}$ was found to have a metal-to-insulator (MI) phase
transition near T$_{MI}$ of $\sim$62~K.  \cite{Sekine97} Magnetic
susceptibility and x-ray diffraction measurements \cite{Sekine97} by Sekine
$et$ $al.$ suggested that this MI transition is neither magnetic nor
crystallographic in origin.  Although valence fluctuations of the Pr ion might
lead to this MI transition, a detailed X-ray absorption near edge structure
(XANES) study of the Pr L$_{II}$-edge for PrRu$_4$P$_{12}$ suggested that the
Pr valence stays at 3$^+$ above and below T$_{MI}$. \cite{Lee99} Recent
electron diffraction experiments by Lee $et$ $al.$ showed weak superlattice
spots at 12 and 40~K, which were not observed at 70~K.  \cite{Lee01} The
observation was then attributed to a structural phase transition from
$Im\bar{3}$ to another cubic phase, $Pm\bar{3}$, as temperature decreases from
70 to 40~K. In the $Pm\bar{3}$ phase, the Ru atoms are displaced along (1,1,1)
directions.  It was also suggested theoretically by Curnoe $et$ $al.$ that a
symmetry lowering (1,1,-2) Ru displacement leading to a $Pmmm$ phase may be
responsible for this MI transition, though a (1,1,1) distortion mode leading to
a $Pm\bar{3}$ phase could not be excluded.  \cite{Curnoe02,Curnoe02a} Band
structure calculations also showed that a very tiny P distortion could open a
band gap at the Fermi surface, leading to the MI phase transition in
PrRu$_4$P$_{12}$. \cite{Harima03}

Despite many experimental and theoretical studies of PrRu$_4$P$_{12}$
indicating that a structural distortion is responsible for the MI transition,
there is still a lack of clear experimental evidence, which hinders a deeper
understanding of this phenomenon. The main problem is that any structural
distortion associated with the MI transition has to be very small, so that
typical diffraction techniques cannot definitively observe it. Although recent
synchrotron x-ray diffraction experiment on a single crystal PrRu$_4$P$_{12}$
suggested a $Pm\bar{3}$ phase at 10~K, \cite{Lee04} no high-T data were
reported for comparison purposes. The extended x-ray absorption fine structure
(EXAFS) technique, however, is ideally suited to detect small changes in local
structure.  We, therefore, carried out EXAFS experiments on both
PrRu$_4$P$_{12}$ and PrOs$_4$P$_{12}$, because PrOs$_4$P$_{12}$ is a close
structural relative of PrRu$_4$P$_{12}$, but is a metal below room temperature.
Our data clearly show that a structural distortion of an unexpected nature
accompanies the MI transition in PrRu$_4$P$_{12}$.

EXAFS data at the Pr $L_{III}$-, Os $L_{III}$- and Ru $K$-edges were collected
as a function of temperature at beam line 20-BM at the Advanced Photon Source
(APS).  Single crystal samples were ground into very fine powders and brushed
onto scotch tape for the measurements (particle size $\sim$ 10 $\mu$m).
Energy-space data were reduced using standard procedures
\cite{Li95b,Bridges95b} and the resulting $k$-space data were Fourier
transformed to $r$-space yielding peaks that correspond to different atomic
shells.  By fitting each $r$-space peak to a theoretical function calculated
using the FEFF8 program \cite{FEFF8} plus a Gaussian pair distribution
function, we extracted the peak widths $\sigma$ and peak distances $r$.
$\sigma$ provides quantitative information about the local distortions,
including thermal vibration and static distortion, of a particular atomic
shell. 

Fits to the Pr $L_{III}$-edge $r$-space data were performed for the first two
atomic shells (Pr-P and Pr-Ru(Os)). Panel (a) of Fig. \ref{pr_data} shows the
fit result to the 25~K data for PrRu$_4$P$_{12}$ as an example of the fit
quality. Since the filler atoms stay in a relatively large cage in the filled
skutterudites, it is possible that the filler atoms might be displaced to an
off-center position. Indeed, for PrOs$_4$Sb$_{12}$, the larger value of
$\sigma^2$ at low temperatures indicated additional distortions which could be
caused by a small Pr off-center displacement.\cite{Cao03} We, therefore,
suspected that the MI transition in PrRu$_4$P$_{12}$ might be due to a Pr
off-center displacement at low temperatures. However, $\sigma^2$ for the Pr-P
bonds, extracted from the fits (panel (b) of Fig.  \ref{pr_data}) for
PrRu$_4$P$_{12}$ and PrOs$_4$P$_{12}$ show no obvious anomalies as the
temperature drops below 60~K. The rattling of the filler atoms is considered to
be a localized vibration and, therefore, should be consistent with an Einstein
model. \cite{Keppens98} As shown in panel (b) of Fig. \ref{pr_data}, both sets
of data can be very well fit to an Einstein model with no obvious deviation
across T$_{MI}$. In fact, the static distortions, reduced masses and Einstein
temperatures obtained in PrRu$_4$P$_{12}$ and PrOs$_4$P$_{12}$ are very
similar, which suggests that the Pr local structure is essentially the same in
both samples (static distortions $\sim$0.00003(5) {\AA}$^2$; reduced masses
$\sim$112(5) g/mol; Einstein temperatures $\sim$130(3)~K). Thus, we have no
evidence for Pr off-center displacement in PrRu$_4$P$_{12}$ at any temperature,
and consequently conclude that Pr displacement cannot be the origin of the MI
transition in this material.

We now compare the Ru $K$-edge data and the Os $L_{III}$-edge data for the two
materials. Fig. \ref{rspace_fit} plots the $r$-space data and the fit result at
the lowest temperatures for each edge.  The four nearest atomic shells
(Ru(Os)-P, Ru(Os)-Pr, second Ru(Os)-P, and Ru(Os)-Ru(Os)) were used to fit the
data. To account for a possible interference between further multi-scattering
peaks and the Ru(Os)-Ru(Os) peak, two additional multi-scattering peaks at
longer distances with small amplitudes were also included in the fit. 

Fig. \ref{ruos_s2} shows the temperature dependence of $\sigma^2$ together with
fits of $\sigma^2$(T) to a Debye model for the Ru(Os)-P and Ru(Os)-Ru(Os) atom
pairs. For the Ru-P and Os-P bonds, no anomaly in $\sigma^2$(T) was found
across T$_{MI}$. The high Debye temperatures obtained for these two bonds
($\Theta_D$ = 690~K for Ru-P and 720~K for Os-P) suggest that the Ru-P and Os-P
bonds are rather strong, making the RuP$_6$ and OsP$_6$ octahedra in
PrRu$_4$P$_{12}$ and PrOs$_4$P$_{12}$ rigid units.  The most interesting
feature in this figure is that the $\sigma^2$ of the Ru-Ru pair starts to
increase as T drops below 60~K ($\approx$T$_{MI}$).  If thermal vibrations
dominated the broadening, $\sigma^2$ should gradually decrease as T decreases,
saturating at the zero-point-motion value.  The $\sigma^2$ for the Os-Os atom
pair indeed follows this trend ($\Theta_D\approx$370~K).However, the increase
in $\sigma^2$ for the Ru-Ru pair at low temperatures clearly indicates that an
extra distortion other than thermal vibration exists below 60~K in
PrRu$_4$P$_{12}$. Fitting the Ru-Ru $\sigma^2$ data from 70 to 300~K yields
$\Theta_D$ of $\sim$ 398~K with zero static distortion (the solid curve in Fig.
\ref{ruos_s2}). Despite the excellent agreement with the T$\geq$70~K data, this
fit yields a curve distinctly below the data for T$\leq$60~K.  Attempts to fit
the Ru-Ru data at all temperatures to a single Debye model plus a small static
distortion (the dotted curve in Fig.  \ref{ruos_s2}) yields a $\Theta_D$ of
413~K and static distortion of 0.00018 {\AA}$^2$. This model, however, fails to
give a good fit to either the high- or low-temperature data.  Thus there must
be an extra Ru-Ru distortion at temperatures below 60~K.

We now discuss the origin of this Ru-Ru distortion. In the undistorted lattice,
all Ru-Ru distances are the same, so that any small expansion or contraction of
the Ru framework that does not break this degeneracy cannot lead to a larger
$\sigma^2$ for the Ru-Ru peak. Thus, this extra distortion found for the Ru-Ru
pair at low temperatures must come from a Ru displacement that breaks the
degeneracy of the Ru-Ru distance.  Lee $et$ $al.$ proposed that a $Im\bar{3}$
to $Pm\bar{3}$ phase transition is the origin of the MI transition in
PrRu$_4$P$_{12}$. \cite{Lee01} Due to a (1,1,1) Ru distortion, the Ru
sublattices then have two different sizes in the $Pm\bar{3}$ phase. This splits
the six nearest Ru-Ru atom pairs in the $Im\bar{3}$ phase into 3 long and 3
short pairs, which could produce an extra broadening of the Ru-Ru peak in the
EXAFS $r$-space spectra. Although the low-T $Pmmm$ phase suggested by Curnoe
$et$ $al.$ changes the cubic Ru sublattice to tetragonal and also creates
Ru-Ru distortion, \cite{Curnoe02,Curnoe02a} recent inelastic neutron
scattering measurements show a $\Gamma_1$ ground state for Pr$^{3+}$ in
PrRu$_4$P$_{12}$, \cite{Iwasa04} indicating that the assumption of a $\Gamma_3$
ground state in Curnoe's theory is not correct. Furthermore, the number of
transitions of excited crystal field levels observed in the inelastic neutron
scattering experiments below T$_{MI}$ are consistent with the two inequivalent
Pr sites in the $Pm\bar{3}$ phase. Therefore, the possibility of
the low-T $Pmmm$ phase can be excluded.

To test the $Pm\bar{3}$ model, we then fit the Ru-Ru peak in the EXAFS data
with two Ru-Ru peaks. Note that EXAFS cannot normally resolve two peaks split
by a small distance, as the fitting quality will be comparable for both a
single-peak and a two-peak fit. However we can test whether the excess
broadening observed at low T is consistent with the scenario suggested in the
$Pm\bar{3}$ phase under the assumption that the random static distortions are
small. In this new fit, the number of neighbors was again fixed and the
amplitude ratio of the two Ru-Ru peaks was set to be 1:1, each with degeneracy
of three. Also, the $\sigma$ parameters for these two peaks were set to be the
same, because the difference in the pair distances must be very small. Thus,
only the Ru-Ru atom distances were allowed to change independently. As shown in
panel (a) of Fig.  \ref{2peak_fit}, the low-temperature $\sigma^2$ from the
double-peak model are now lowered and very consistent with the $\Theta_D$=398~K
Debye model which was used to fit the data from 70 to 300~K in the single peak
model. The Ru-Ru peak distances obtained from the two fits are also plotted in
Fig. \ref{2peak_fit} (panel (b)). The double-peak fit displays two Ru-Ru
distances below $T_{MI}$, which are separated by about 0.022$\sim$0.030 {\AA}.
This value is consistent with recently obtained low-temperature synchrotron
x-ray diffraction results.  \cite{Lee04} The result of the double-peak fit is
clearly consistent with a $Pm\bar{3}$ space group for T$\leq{T_{MI}}$.  Above
$T_{MI}$, the two Ru-Ru peaks collapse back together, confirming a
high-temperature $Im\bar{3}$ phase.  

As pointed out earlier, because the RuP$_6$ octahedra are quite rigid units
forming a corner-shared framework, a slight rotation, plus a tiny displacement,
of the RuP$_6$ octahedra is needed to produce the observed Ru displacement.
This inevitably leads to tiny correlated P displacements. However, because
these P displacements are not principally along the radial directions of the
Pr-P bonds, the subsequent Pr-P bond-length variations are too small to be
observed in EXAFS (Panel (b) of Fig.  \ref{pr_data}).
                                                                                
Our EXAFS data clearly show that the MI transition in PrRu$_4$P$_{12}$ is
associated with small correlated Ru and P displacements. Such atomic
displacements may open a band gap at the Fermi surface, leading to
semiconducting (insulating) behavior at low temperatures.  A band structure
calculation by Harima $et$ $al.$ indeed showed that very small P displacements
can introduce a band gap in PrRu$_4$P$_{12}$.  \cite{Harima03} However,
possible Ru displacements were not considered in that work. Thus, future
theoretical calculations need to consider correlated displacements of Ru
and P caused by a rotation/displacement of the RuP$_6$ octahedra.

We also note that recent inelastic neutron scattering experiments of
PrRu$_4$P$_{12}$ have been interpreted \cite{Iwasa04} as evidence for a
charge-density-wave produced by strong 4f-hybridization. Our results may help
to illuminate this issue.  Such a local-distortion-induced MI transition in
PrRu$_4$P$_{12}$ may also be related to the semiconducting behavior observed in
other phosphide skutterudites, such as CeRu$_4$P$_{12}$ and CeFe$_4$P$_{12}$.
In this regard, the estimated static contribution to $\sigma^2$ for the Ru-Ru
atom pair from the Ru $K$-edge data for CeRu$_4$P$_{12}$ is about 0.00093
{\AA}$^2$ at all temperatures;\cite{Cao04} assuming a distortion of the Ru
positions, this corresponds to a Ru displacement $\sim$0.03 {\AA}
($\sqrt{\sigma^2_{static}}$).  A similar static distortion for the Fe-Fe atom
pair in CeFe$_4$P$_{12}$ was also inferred. 

In summary, we present the first comprehensive local structure measurements of
PrRu$_4$P$_{12}$. We find that a static Ru displacement is clearly associated
with the MI transition in PrRu$_4$P$_{12}$, while the P displacements must be
associated with the motion of the rigid RuP$_6$ octahedra. This Ru displacement
found in PrRu$_4$P$_{12}$ suggests that the low temperature phase in this
material is likely to be $Pm\bar{3}$.  This EXAFS experiment clearly
illustrates the importance of correlating the effects of local distortions with
the transport properties of the filled skutterudites.

\begin{acknowledgements}
This work was carried out under the auspices of the U.S. Department of Energy,
Office of Science. The experiments were preformed at APS, which is funded by
the DOE, Office of Sciences, Office of Basic Energy Sciences. Research at UCSD
was supported by the US Department of Energy under Grant No. FG02-04ER46105.
\end{acknowledgements}


\begin{thebibliography}{10}
                                                                                
\bibitem{Meisner85}
G.~P. Meisner $et$ $al$, J. Appl. Phys. {\bf 57},  3073  (1985).
                                                                                
\bibitem{Shirotani99}
I. Shirotani $et$ $al$, J. Solid State Chemistry {\bf 142},  146  (1999).
                                                                                
\bibitem{Sekine97}
C. Sekine $et$ $al$, Phys. Rev. Lett. {\bf 79}, 3218  (1997).
                                                                                
\bibitem{Lee99}
C.~H. Lee $et$ $al$, Phys. Rev. B {\bf 60},  13253  (1999).
                                                                                
\bibitem{Lee01}
C.~H. Lee $et$ $al$, J. Phys.: Condens. Matter {\bf 13},  L45 (2001).
                                                                                
\bibitem{Curnoe02}
S.~H. Curnoe $et$ $al$, Physica B {\bf 312-313},
  837  (2002).
                                                                                
\bibitem{Curnoe02a}
S.~H. Curnoe $et$ $al$, J. of Phys. Chem. Solids {\bf 63},  1207  (2002).
                                                                                
\bibitem{Harima03}
H. Harima $et$ $al$, Acta Physica Polonica B {\bf 34},  1189  (2003).
                                                                                
\bibitem{Lee04}
C. H. Lee $et$ $al$, J. Magn. Magn. Mater. {\bf 272-276}, 426  (2004).
                                                                                
\bibitem{Li95b}
G.~G. Li, F. Bridges, and C.~H. Booth, Phys. Rev. B {\bf 52},  6332  (1995).
                                                                                
\bibitem{Bridges95b}
F. Bridges, C.~H. Booth, and G.~G. Li, Physica B {\bf 208\&209},  121  (1995).
                                                                                
\bibitem{FEFF8}
A.~L. Ankudinov $et$ $al$, Phys. Rev. B {\bf 58},  7565  (1998).
                                                                                
\bibitem{Cao03}
D. Cao $et$ $al$, Phys. Rev. B {\bf 67},  180511(R) (2003).
                                                                                
\bibitem{Keppens98}
V. Keppens $et$ $al$, Nature {\bf 395},  876  (1998).
                                                                                
\bibitem{Iwasa04}
K. Iwasa $et$ $al$, submitted to Phys. Rev. Lett.  (2004).
                                                                                
\bibitem{Cao04}
D. Cao $et$ $al$, Phys. Rev. B, In press  (2004).
                                                                                
\end{thebibliography}

\begin{figure}
\caption{Panel (a): Pr $L_{III}$-edge $r$-space data for PrRu$_4$P$_{12}$ at
25~K and the fit result. The Fourier transform range is from 3.0 to
10.5 {\AA}$^{-1}$, with 0.3 {\AA}$^{-1}$ Gaussian broadening. The fit range
is from 2.3 to 3.7 {\AA}. The high
frequency curve inside the envelope is the real part of the Fourier Transform
(FT$_R$). The envelope is defined as: $\protect\pm\sqrt{FT_R^2 + FT_I^2}$,
where FT$_I$ is the imaginary part of the Transform. There is a well-defined
EXAFS phase shift for each peak, consequently the nearest Pr-P peak occurs
at $\sim$2.7 {\AA} and the Pr-Ru peak is at $\sim$3.4 {\AA}. The inset displays
Pr $L_{III}$-edge $k$-space data at 25~K as an example to show the quality of
the EXAFS data. Panel (b): the temperature dependent $\sigma^2$ obtained from
fits for the Pr-P bond in PrRu$_4$P$_{12}$ and PrOs$_4$P$_{12}$
samples. The results of fitting $\sigma^2$ $vs.$ T to an Einstein
model for the each sample are also plotted.}
\label{pr_data}
\end{figure}

\begin{figure}
\caption{A plot of the Ru $K$- and Os $L_{III}$-edges $r$-space data at 28 and
25~K along with the fit result for PrRu$_4$P$_{12}$ and PrOs$_4$P$_{12}$,
respectively. The Fourier transform range is from 3.0 to
15.0 {\AA}$^{-1}$, with 0.3 {\AA}$^{-1}$ Gaussian broadening for both edges.
The fit range is from 1.2 to 4.0 {\AA} for the Ru $K$-edge data; while it is
1.2 to 4.1 {\AA} for the Os $L_{III}$-edge data. The Ru-P peak is at $\sim$1.8
{\AA} and the Ru-Ru peak is at $\sim$3.8 {\AA}; while the Os-P peak is at
$\sim$2.0 {\AA} and the Os-Os peak is at $\sim$3.9 {\AA}. The insets in both
panels plot the $k$-space data for each edge, respectively, to show the
quality of the EXAFS data.}
\label{rspace_fit}
\end{figure}

\begin{figure}
\caption{$\sigma^2$ $vs.$ temperature for the Ru(Os)-P, Ru(Os)-Ru(Os) atom
pairs obtained from the Ru $K$- Os $L_{III}$-edges EXAFS data for
PrRu$_4$P$_{12}$ and PrOs$_4$P$_{12}$, respectively. The Debye model fit to
each set of data is also shown in the figure. For the Ru-Ru pair (solid
circles), $\sigma^2$ increases below 60K; here two
Debye models are used: the dotted line is the fit to the data points at all
temperatures and yields $\Theta_D$ of 413~K; the solid line is the fit result
to the data points for temperatures between 70 and 300~K, and gives $\Theta_D$
of 398~K. The error bar for each data point, which is not shown, is smaller
than the symbol size, except for one data point indicated in the figure. The
error for the Debye temperature is $\sim$20~K.}
\label{ruos_s2}
\end{figure}

\begin{figure}
\caption{Panel (a): $\sigma^2$ of the Ru-Ru atom pair obtained from the
original single Ru-Ru peak fit and the double Ru-Ru peak fit as well as the
$\Theta_D$=398~K Debye model. The inset shows the difference of these
two sets of $\sigma^2$ up to 100~K. Panel (b): The Ru-Ru peak distances
obtained from the single-peak and double-peak fits.
The error bars for all $\sigma^2$ data and
the peak distance in the single-peak fit are smaller than the symbol size and,
therefore, not shown.}
\label{2peak_fit}
\end{figure}

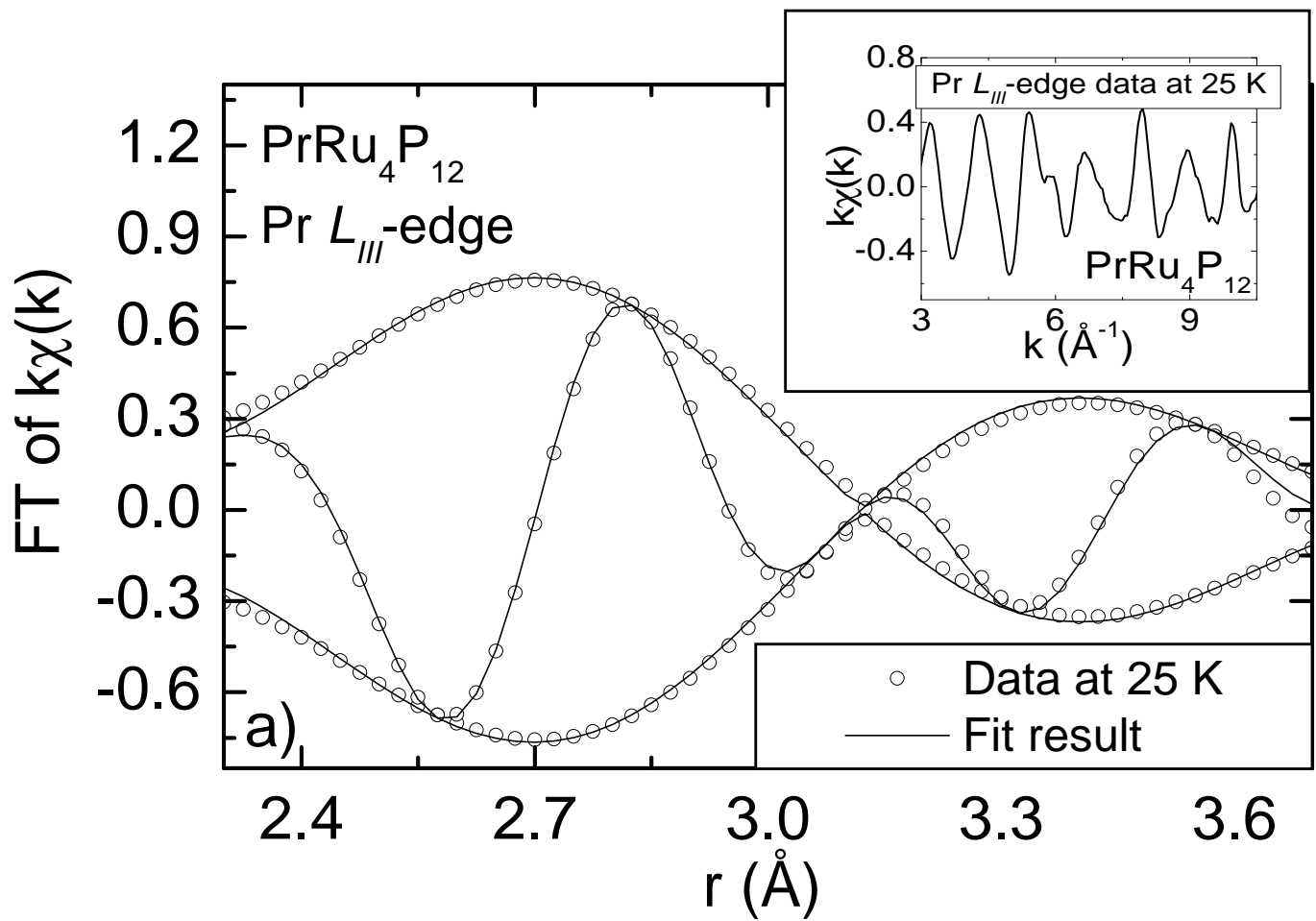
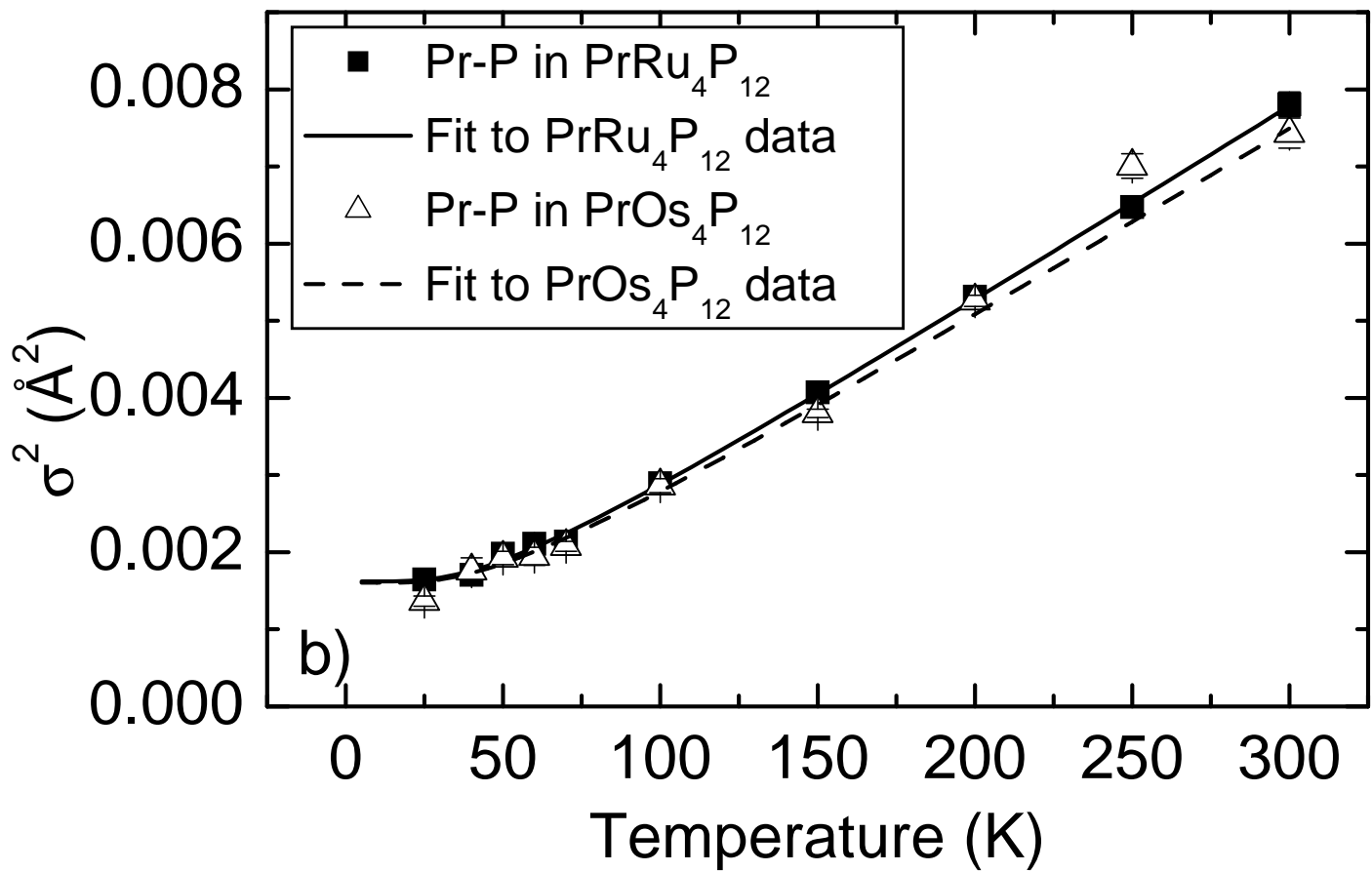
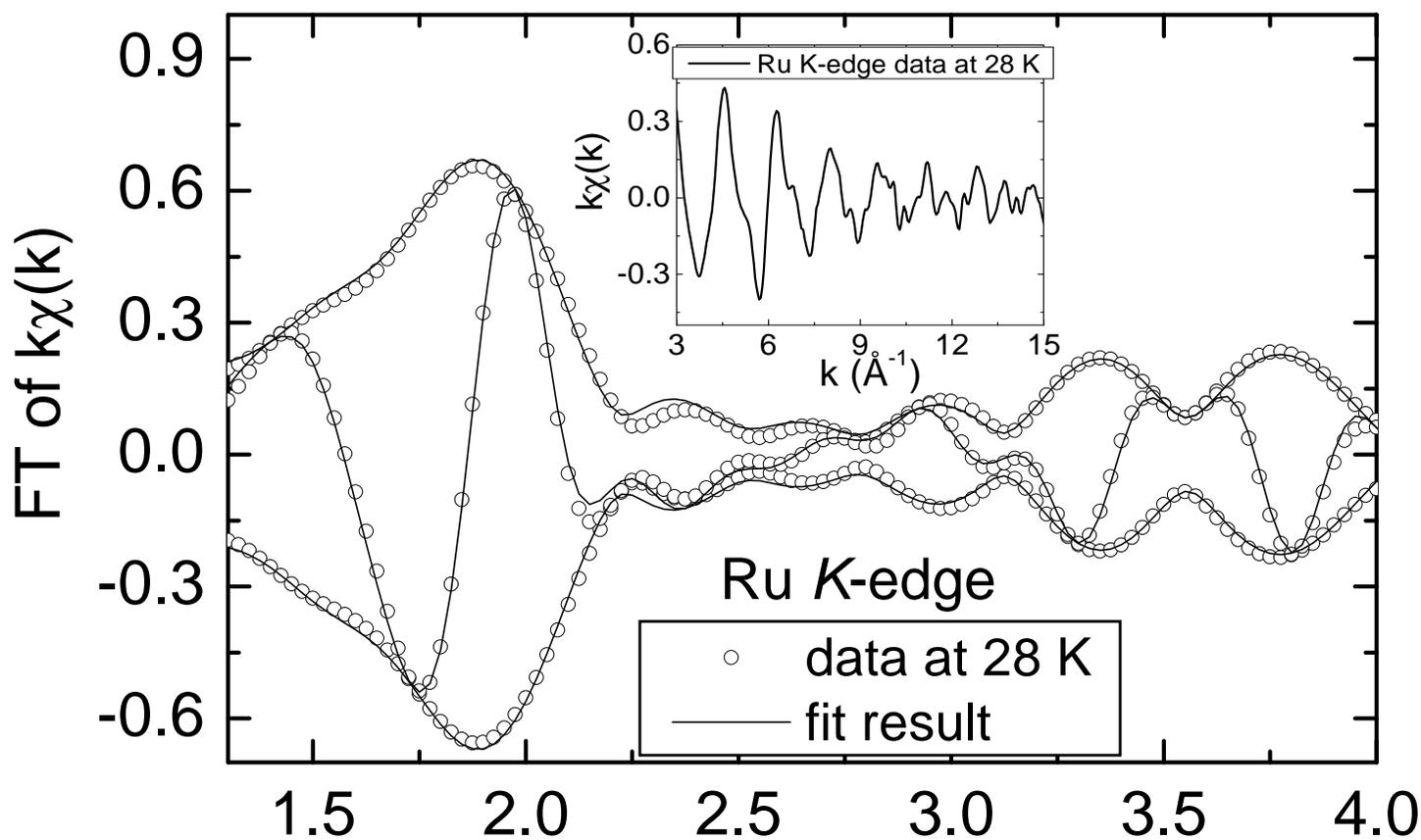
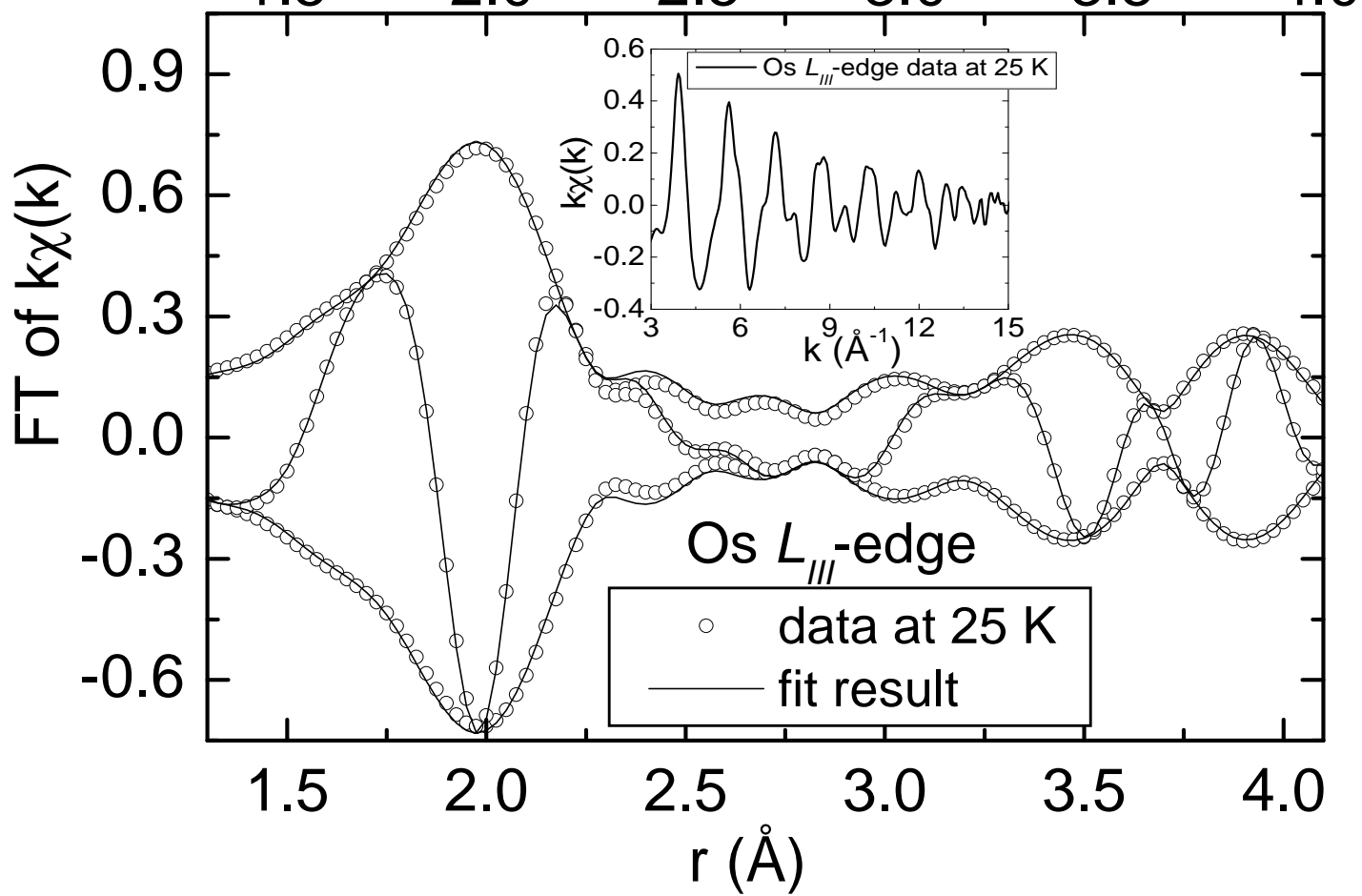
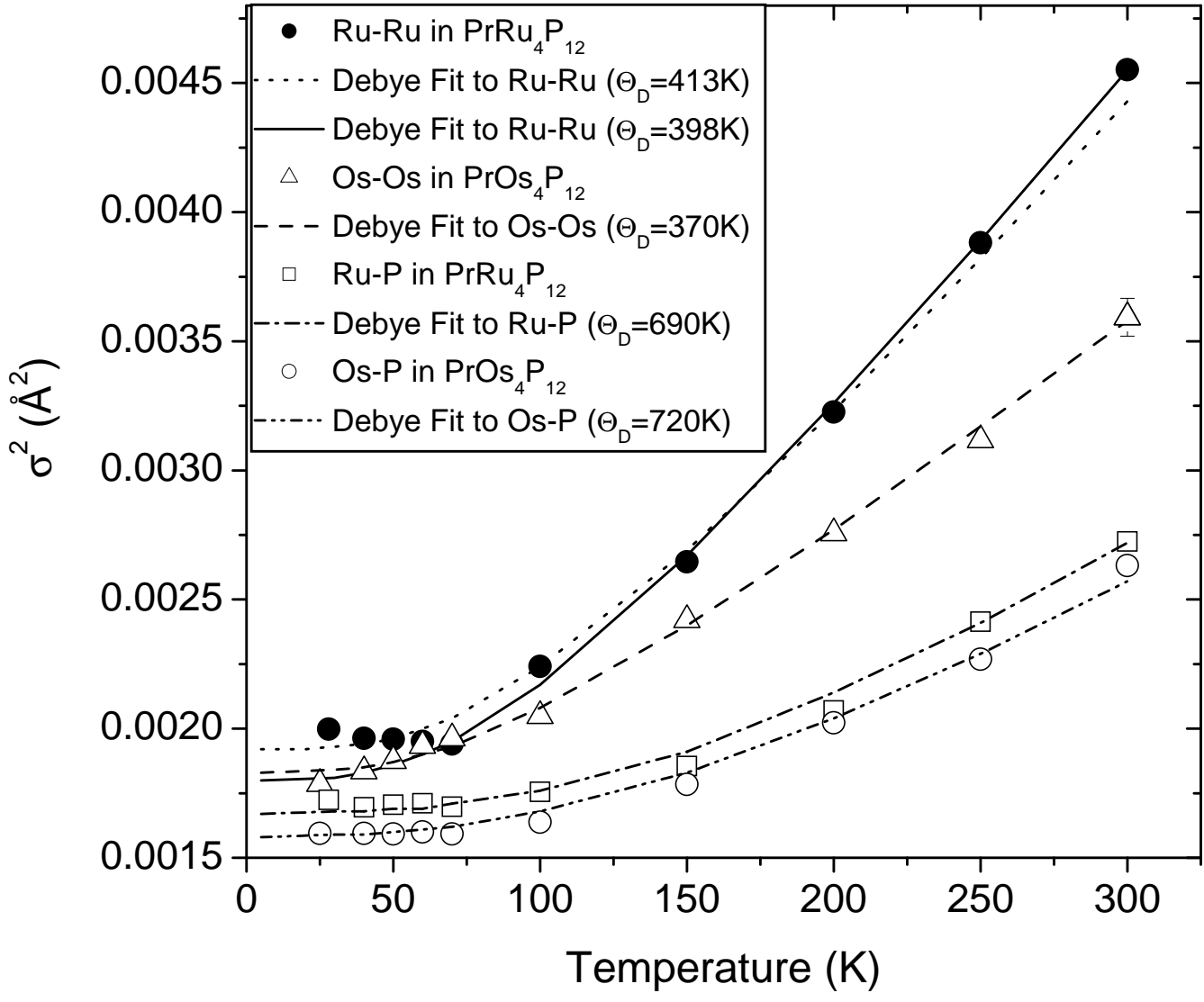
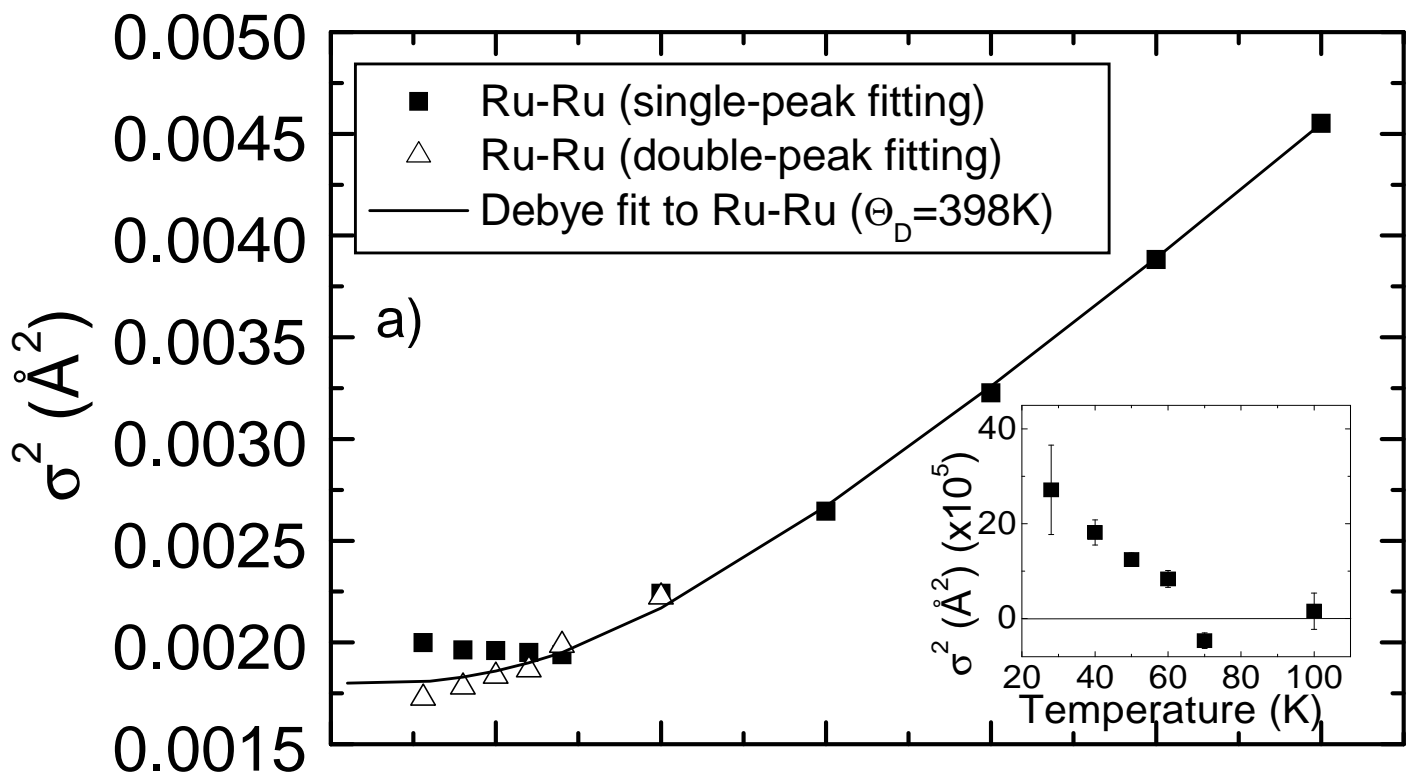
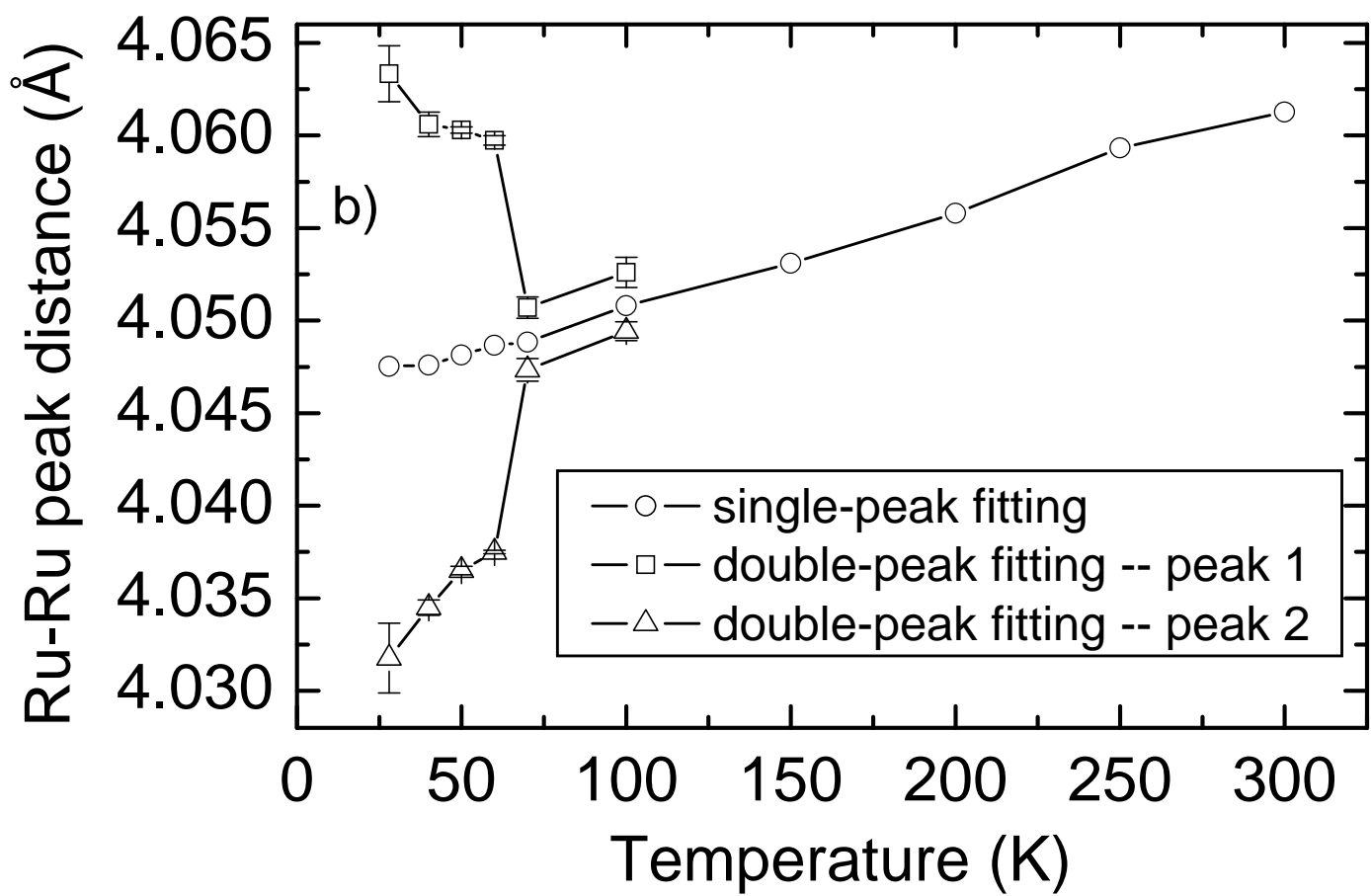
\end{document}